\documentclass[a4paper,11pt]{article}
\pdfoutput=1 % if your are submitting a pdflatex (i.e. if you have
\usepackage{jinstpub} % for details on the use of the package, please
\usepackage{soul}

\usepackage{hyperref}
\usepackage{cleveref}

\usepackage{lineno}
\usepackage{multirow}
\usepackage{caption}
\usepackage{subcaption}
\usepackage{xspace}
\usepackage{eqnarray,amsmath}
\usepackage{indentfirst}
\usepackage{siunitx}

\usepackage{adjustbox}

\usepackage[colorinlistoftodos]{todonotes}
\usepackage{xcolor}

\setstcolor{red}

\newcommand{\sonoarrivatoqui}[1]{\textcolor{red}{\\ \%\%\% sono arrivato qui \%\%\% \\\\}}
\newcommand{\gluedcubelayer}{glued-cube layer\xspace} 
\newcommand{\mppcpcb}{MPPC-PCB\xspace}

\title{Design and characterization of large-size monolithic layers of 3D-segmented plastic scintillator}

\author[a]{J.~Berisha,}
\author[a]{X.~Y.~Zhao,}
\author[b]{A.~Boyarintsev,}
\author[c]{S.~Cap,}
\author[a]{D.~Di Calafiori,}
\author[b]{M.~Kiktev,}
\author[a]{U.~Kose,}
\author[b]{A.~Kuzmina,}
\author[b]{N.~Leskina,}
\author[a]{A.~Rubbia,}
\author[b]{T.~Sibilieva,}
\author[a]{T.~Weber}
\author[a,1]{and D.~Sgalaberna}

\affiliation[a]{ETH Zurich, Institute for Particle Physics and Astrophysics, CH-8093 Zurich, Switzerland}
\affiliation[b]{Institute for Scintillation Materials NAS of Ukraine (ISMA), Kharkiv 61072, Ukraine}
\affiliation[c]{University of Geneva, Section de Physique, DPNC, 1205 Geneva 24, Switzerland}

\emailAdd{davide.sgalaberna@cern.ch}

\note{Corresponding author}

\abstract{
Three-dimensional fine-grained plastic scintillator detectors are extensively used in high-energy physics experiments where particle tracking, identification, calorimetry and sub-nanosecond timing information are required. 
Moreover, the low cost of plastic scintillator makes, in principle, the scalability to multi-tonne detectors possible.
In order to circumvent the difficulties related to the long production time and the challenging assembly of millions of tiny independent scintillating voxels, a new technology consisting of gluing a few thousand $9 \times 9 \times 9$ mm$^3$ plastic scintillator cubes in a single large monolithic layer with machining precision, while ensuring an efficient optical separation, was developed and demonstrated. 
In this work we demonstrate that such technology can be reliably extended to a monolithic layer of $48 \times 48$ optically-isolated scintillating cubes, the size required for future ton-scale detectors, fulfilling the requirements of future neutrino detectors and sampling calorimeters.
}

\keywords{Plastic scintillator, particle detector, assembly, 3D granularity, neutrino detector, calorimeter}

\begin{document}
\maketitle
\flushbottom

\section{Introduction}
\label{sec:intro}

Modern neutrino detectors and sampling calorimeters use segmented plastic scintillator as the active medium that measures the energy loss from ionizing particles, serving to identify the particle type and energy.
At the same time, the optical separation 
allows for the reconstruction of multiple tracks in high-multiplicity events.
Notable examples of neutrino detectors can be found in the MINERvA~\cite{minerva}, MINOS~\cite{minos} and
T2K~\cite{t2k,t2k-fgd} experiments, made of long thin scintillating bars coated with white paint alternated along different directions to provide independent views of the particle interaction.
A similar concept is adopted by the T2K ND280 sampling calorimeters ~\cite{t2k-ecal}.

An additional step is provided by the full three-dimensional (3D) granularity of the scintillator.
For instance, sampling calorimeters in Ref.~\cite{calice} are made of scintillating tiles directly coupled to silicon photomultipliers (SiPMs). 
In neutrino experiments, examples of scintillator detectors with 3D granularity include the one deployed at the SoLid reactor experiment~\cite{solid} and the Super Fine-Grained Detector (SuperFGD) \cite{superfgd,T2K:2026zms} at the T2K long-baseline oscillation experiment.
The latter is made of almost \num{2000000} 1 cm$^3$ scintillating cubes, each one read out by three orthogonal WLS fibers, for a total of \num{56000} SiPMs.
Such a configuration provides sufficient target mass as well as three orthogonal views for 3D tracking of the multiple particles produced by O(GeV) neutrino interactions.
Each cube was made with injection molding, and three orthogonal holes were drilled to host the WLS fibers. The assembly of the scintillating cubes in the mechanical box required special methods necessary for alignment.  
The whole process, described in detail in Ref.~\cite{T2K:2026zms}, was tedious, complex, and time-consuming.
Moreover, assembling a matrix of many millions of single cubes can introduce complexities due to stack-up tolerances that must be carefully taken into account in the detector design and manufacturing. 

A detector similar to SuperFGD has been proposed for the upgrade of the FASER experiment in view of the high-luminosity run at the LHC (HL-LHC) \cite{FASER:2025myb}.
Such a configuration is also among the options under consideration for a possible upgrade of the Hyper-Kamiokande magnetized near detector \cite{Hyper-Kamiokande:2025asb}.
3D printing is also being developed to easily achieve 3D optical segmentation in a monolithic block of scintillator \cite{3det}. Additionally, new concepts that simply avoid segmentation while preserving tracking capability are currently under development \cite{LiquidO:2019mxd,platon}. However, none of the above technologies have been scaled to large volumes yet.

In Ref.~\cite{Boyarintsev:2021uyw}, we reported an alternative method for producing a large number of optically-separated scintillating elements within a single monolithic layer. The design consisted of a layer made of $5 \times 5$ plastic scintillator cubes joined with white epoxy glue with machining precision.
Ten horizontal 5-cm-long grooves were made to host orthogonal WLS fibers, while twenty-five 1-cm-long holes with a diameter of 1.5 mm were drilled to accommodate the WLS fibers on the third orthogonal axis. 
Like SuperFGD, this design provides three orthogonal projections of particle interactions.

As customized scintillator configurations often require special production facilities that can differ between projects,
a flexible industrial solution based on standard technology, which does not require a special production area, was implemented at ISMA to enable both the manufacturing of single prototypes and mass production.
In this work, we demonstrate the feasibility of this scintillator technology for large detector volumes and its maturity for high-energy physics experiments, setting the basis for the construction of future massive 3D fine-granularity plastic scintillator detectors. 
The specific size of the layer, $48 \times 48$ scintillating cubes glued together, was chosen to match the preliminary detector design for the possible future upgrade of the FASER neutrino detector \cite{FASER:2025myb}.
After describing the design 
in Sec.~\ref{sec:design}, we report in Sec.~\ref{sec:measurements} the evaluated performance in terms of geometrical tolerances, light yield, and optical crosstalk, with measurements carried out at ETH Zurich.

\section{Design and production of the 3D-segmented scintillator layer}
\label{sec:design}

The plastic scintillator used in this work is UPS 923A 
\cite{Artikov:2005mg,Senchishin:2006qw}.
It consists of highly-transparent polystyrene doped by weight with 2\% of PTP and 0.05\% of POPOP.
The scintillator layer is produced with cast polymerization~\cite{cast-polymerization}, where a liquid monomer with dissolved dopants is poured into a mold and heated. After cooling, a rigid solid plastic is obtained.
Cubes with 9 mm edges are cut with a CNC machine, leaving a gap of 1 mm filled with white, reflective epoxy-modified resin.
After curing, the outermost surface of the layer is polished, and $\text{TiO}_2$ paint is uniformly distributed for a maximum thickness of 0.2 mm.
The reflection coefficient at 418 nm for \SI{1}{\milli\meter} of epoxy glue in a \SI{1}{\milli\meter} thick layer is 82\%, while that of a \SI{200}{\micro\meter} thick layer of white paint is 92\%.

\begin{figure}[htbp]
    \centering
    \begin{subfigure}{0.45\textwidth}
        \centering
        \includegraphics[width=\textwidth]{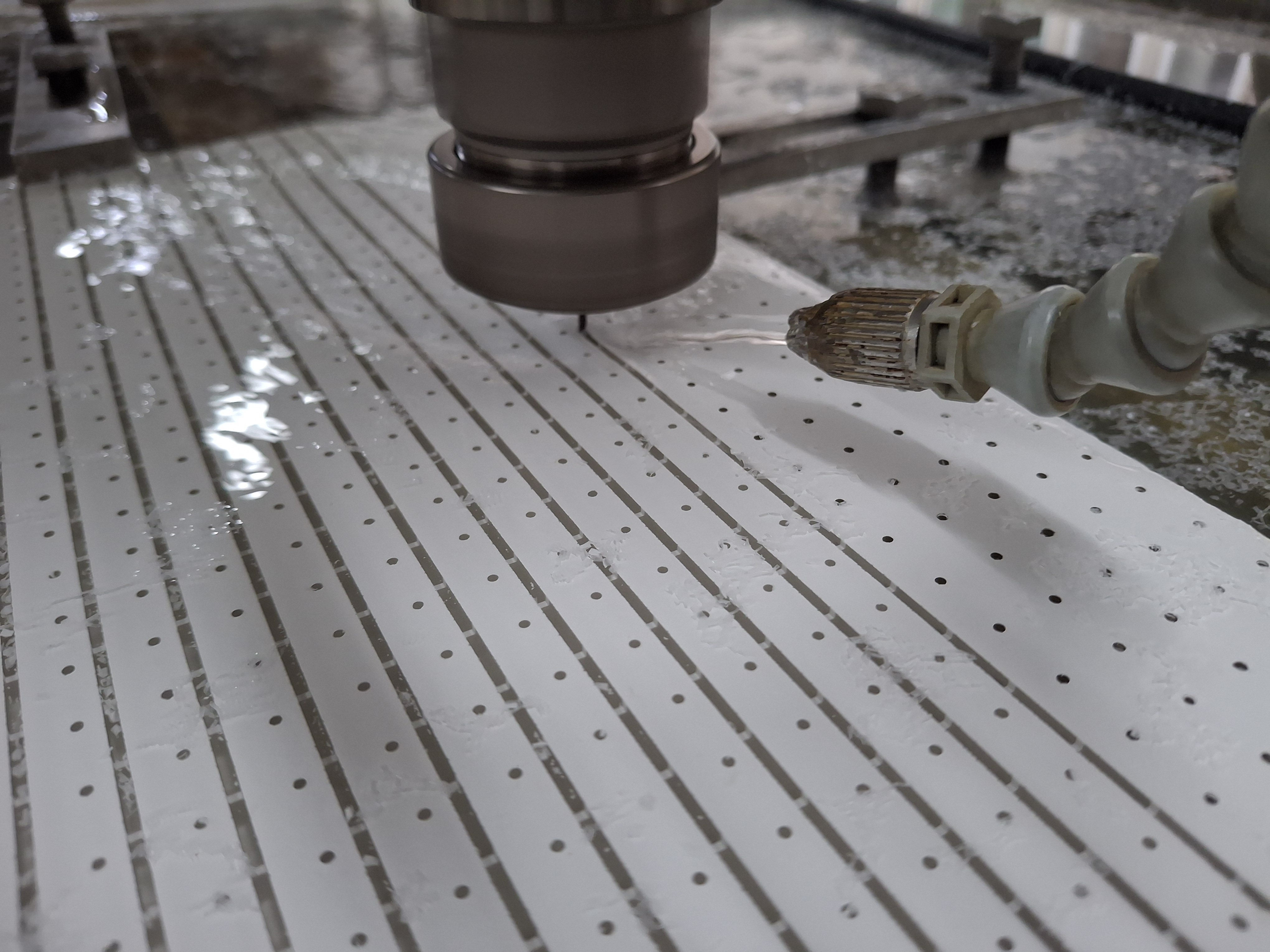}
        \caption{CNC machining of the grooves.}
        \label{fig:large_glued_layer_machining}
    \end{subfigure}
    \hfill
    \begin{subfigure}{0.45\textwidth}
        \centering
        \includegraphics[width=\textwidth]{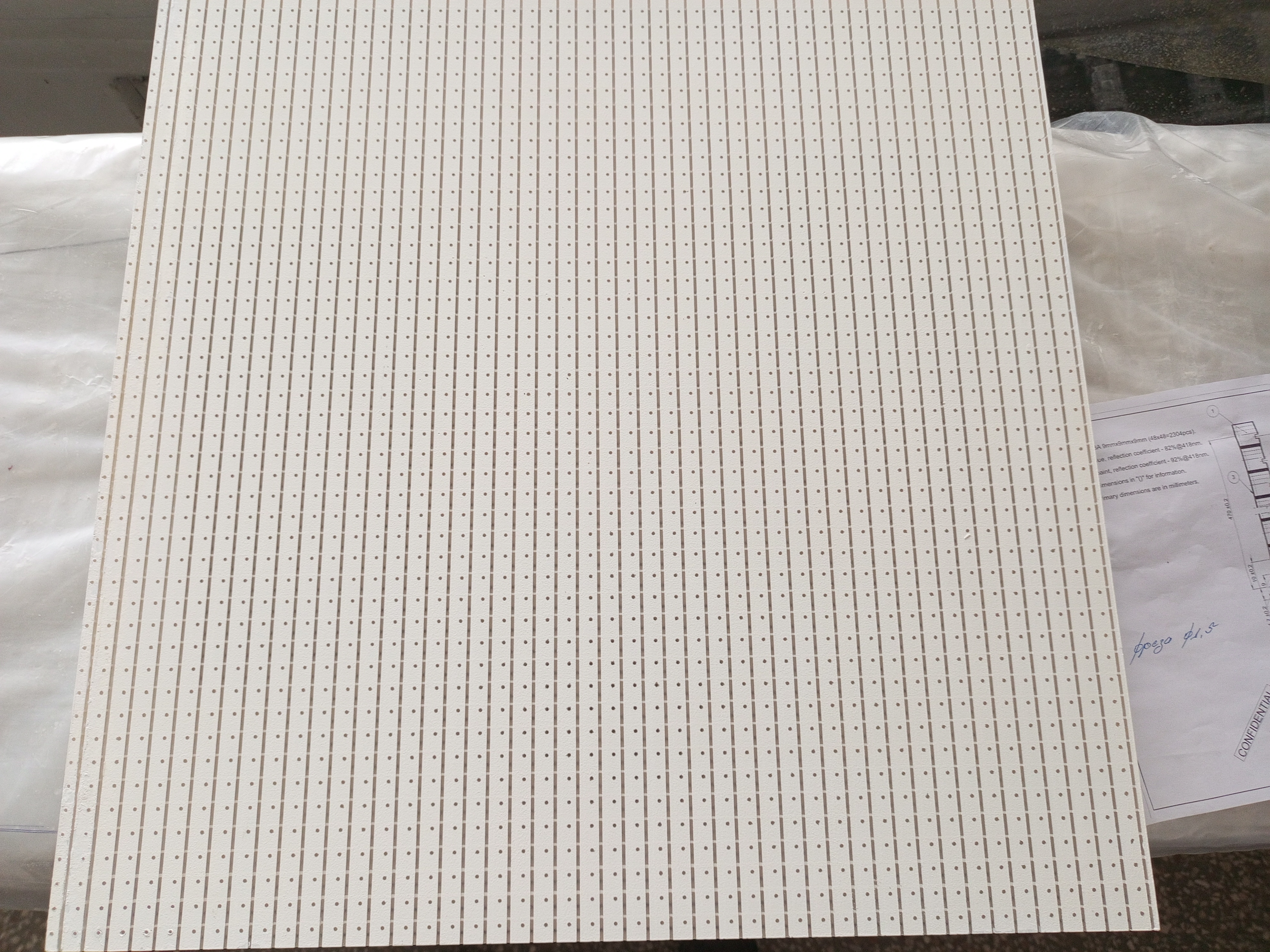}
        \caption{After completion of the manufacturing.}
        \label{fig:large_glued_layer_finished}        
    \end{subfigure}
    \caption{Manufacturing of a \gluedcubelayer.}
    \label{fig:large_glued_layer}
\end{figure}

A 2D matrix of glued cubes is optically separated on all six faces.
The epoxy resin provides mechanical strength and rigidity to the 3D matrix and keeps the scintillation photons trapped within the single cube.
Finally, a CNC machine is used to create the two orthogonal grooves, 1.6 mm wide and 1.5 mm deep, respectively, on the top and bottom faces of the panel, as well as the vertical holes of 1.5 mm in diameter with a 10 mm pitch.
We will refer to it as the \gluedcubelayer.
In Fig.~\ref{fig:large_glued_layer}, the \gluedcubelayer is shown both during the machining of the grooves and after the completion of the manufacturing process. 
The result is a single rigid layer of 2,304 optically-isolated $9\times 9\times 9$~mm$^3$~scintillating cubes organized in a $48 \times 48$ matrix with a 10 mm horizontal pitch, 
including 0.5~mm white epoxy glue on each side,
and a 9.4 mm height, with the 0.2~mm white TiO$^2$ paint on both sides.
Three \gluedcubelayer{s} were manufactured at ISMA.

Although it is possible to produce a multi-layer monolithic block of optically-separated cubes, as shown in Ref.~\cite{Boyarintsev:2021uyw}, it is not possible to read out the scintillation light inside the volume.
In fact, drilling holes in polystyrene becomes difficult even for a thickness of a few centimeters; thus, it is not possible to manufacture the grooves or holes to host the WLS fibers.
The solution is to stack together multiple \gluedcubelayer{s}, alternated with a reflective sheet to ensure 
that no optical crosstalk occurs between the WLS fibers inserted along the orthogonal grooves of the adjacent layers.
In this setup, a specular Mylar sheet covering both sides of each layer was used.
Other types of high-performance reflector sheets, such as the white diffuser lumirror \cite{lumirror,6168236}
or specular mirror \cite{3m_df2000ma,Li:2025dff}, can also be used.
Laser-cut holes allow the vertical fibers to pass through the stack.
The prototype consisting of a stack of three $48 \times 48$ \gluedcubelayer{s} is shown in Fig.~\ref{fig:prototype-setup}.
The three readout views are as follows:
``side view - top groove'' with horizontal WLS fibers on the top side of the \gluedcubelayer;
``side view - bottom groove'' with horizontal WLS fibers on the bottom side of the \gluedcubelayer;
``top view'' with vertical holes for WLS fibers. 

The quality of the manufacturing was evaluated with a coordinate measuring machine with tactile probe
with a maximum uncertainty of 0.1 mm.
The thickness of the \gluedcubelayer was measured at different positions. Its variation was found to be less than 0.2 mm.
The flatness varies within a maximum range of 0.7 mm, with deviations exceeding 0.4~mm observed at the four corners.
This behavior is known in large-format plastic plates and is attributed to the redistribution of internal stresses during the machining process.
Since polystyrene is not a particularly rigid material, its imperfect flatness is not considered a problem.
This was confirmed by tests with multiple layers performed after finalizing the optical measurements.

\section{Performance studies}
\label{sec:measurements}

\begin{figure}
\centering
\raisebox{+0.\height}{
\begin{subfigure}{0.37\textwidth}
\includegraphics[width=\textwidth]{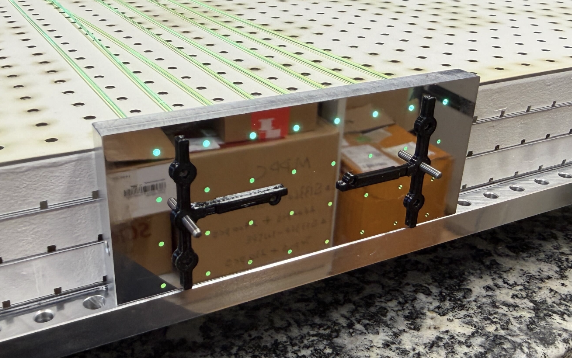}
\caption{Aluminum plate with WLS fibers.}
\label{fig:prototype-alu-plate}
\end{subfigure}
}
\raisebox{+0.\height}{
\begin{subfigure}{0.27\textwidth}
\includegraphics[width=\textwidth]{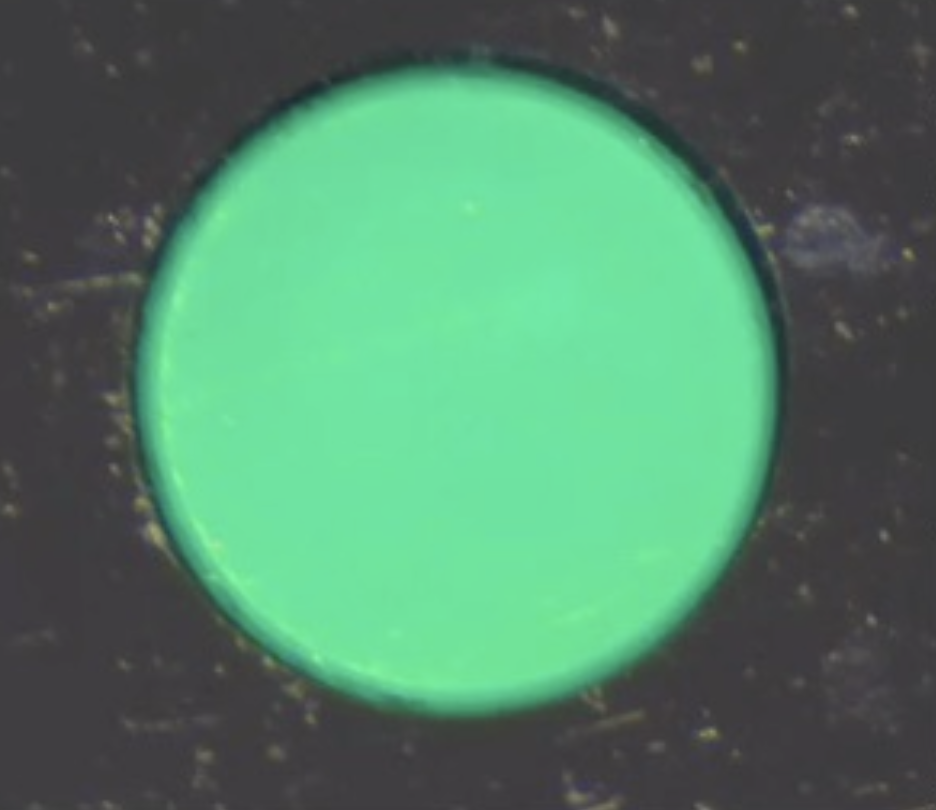}
\caption{Polished WLS fiber.}
\label{fig:prototype-fiber-end}
\end{subfigure}
}
\raisebox{-0.\height}{
\begin{subfigure}{0.3\textwidth}
\includegraphics[width=\textwidth]{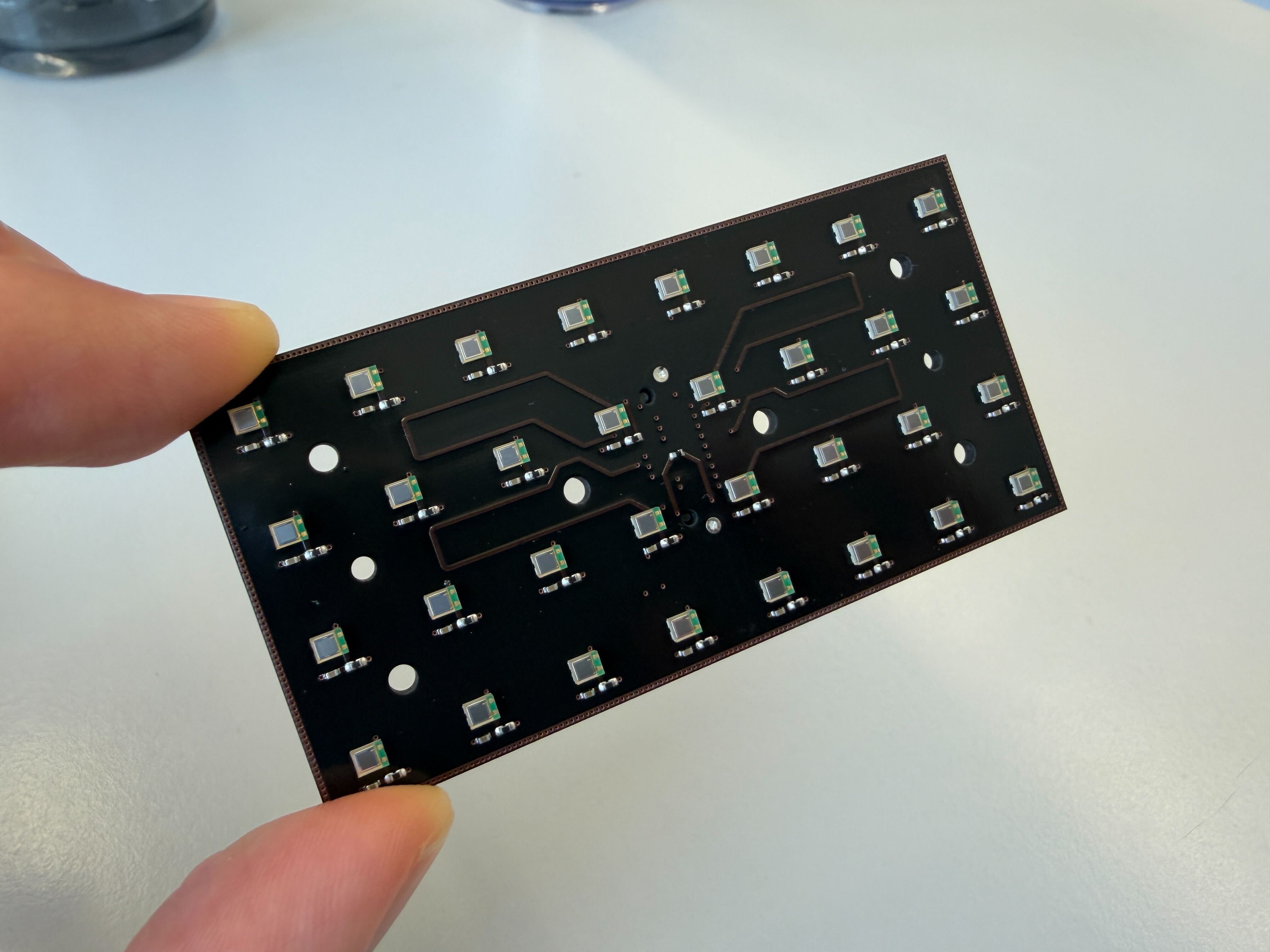}
\caption{\mppcpcb.}
\label{fig:prototype-mppc-pcb}
\end{subfigure}
} 
\\
\raisebox{-0.\height}{
\begin{subfigure}{0.45\textwidth}
\includegraphics[width=\textwidth]{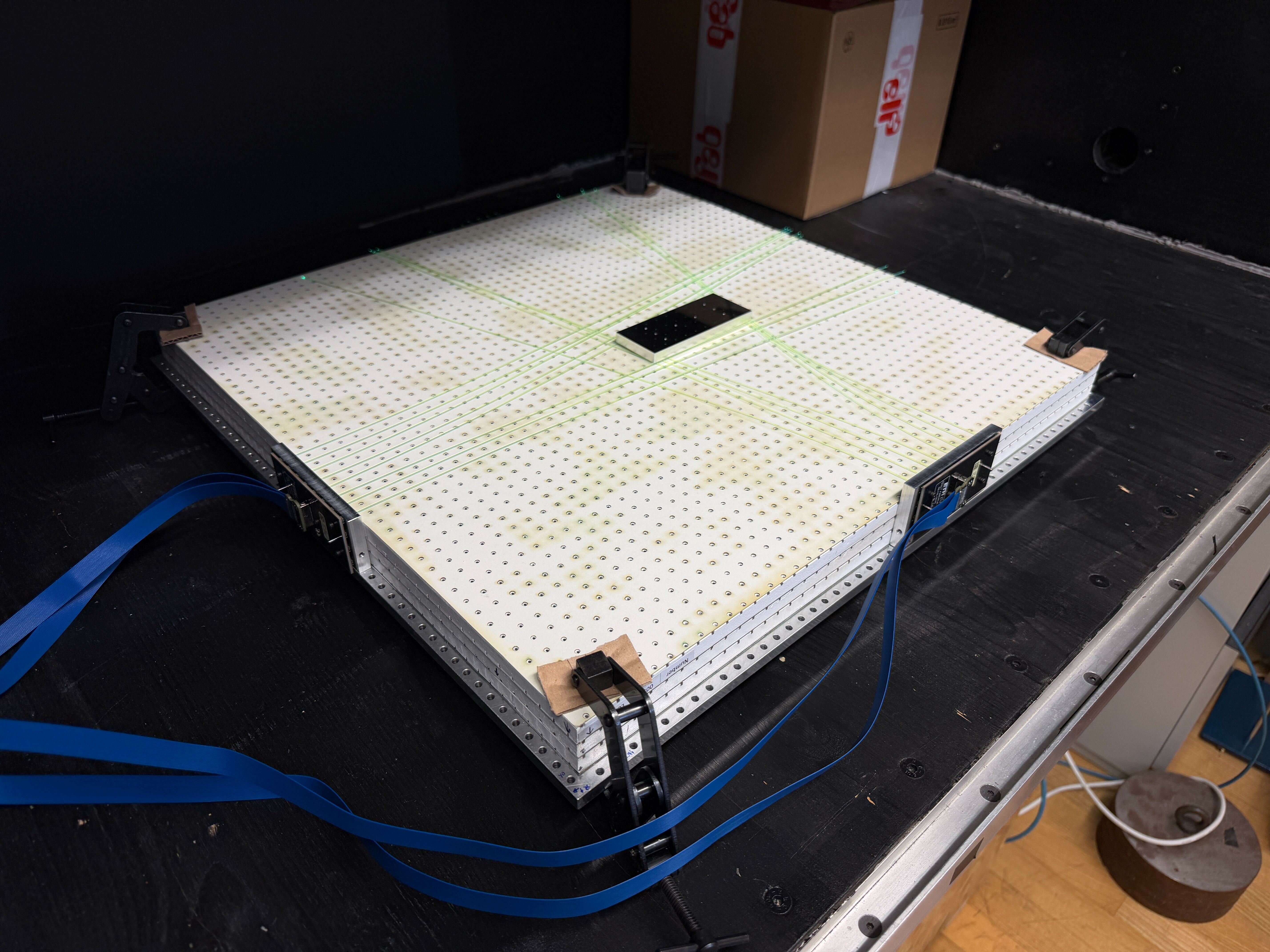}
\caption{Stack of three \gluedcubelayer{s}. On top, the back side of the specular Mylar sheet is visible.}
\label{fig:prototype-setup}
\end{subfigure}
}
\raisebox{-0.\height}{
\begin{subfigure}{0.5\textwidth}
\includegraphics[width=\textwidth]{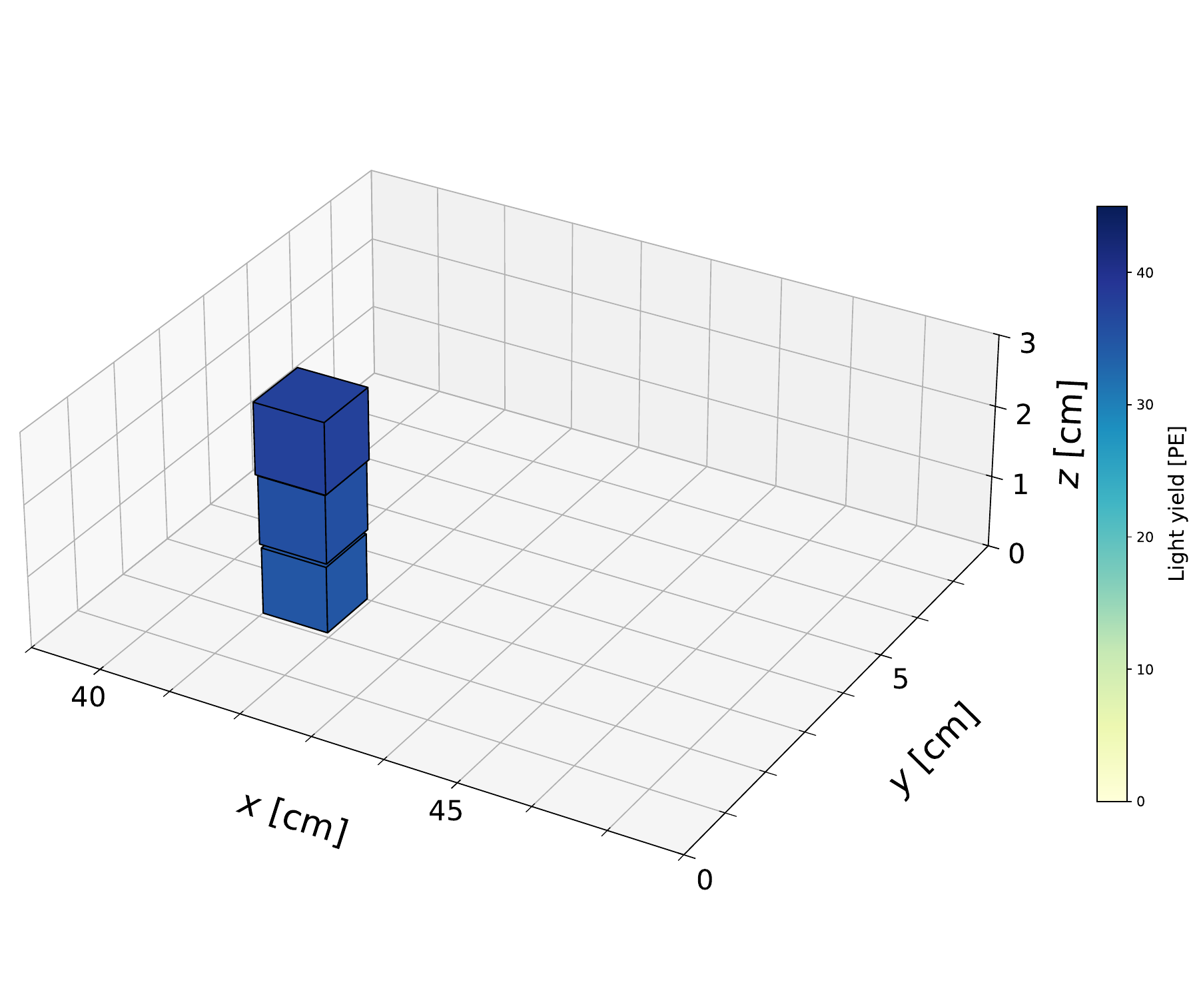}
\caption{A selected cosmic-ray event.
The color shows the number of PE, sum of the two horizontal WLS fibers.}
\label{fig:prototype-event}
\end{subfigure}
}
\caption{
The prototype used to characterize the performance of the \gluedcubelayer and an example of selected cosmic-ray event are shown.
}
\label{fig:prototype}
\end{figure}

In this section, the optical performance of the \gluedcubelayer is reported in terms of scintillation light yield and optical crosstalk between neighboring cubes using a cosmic-ray data sample.
Fig.~\ref{fig:prototype-setup} shows the prototype setup. 
It consists of three \gluedcubelayer{s} stacked together with four bench vises at the corners to fix their positions. Between each pair of layers, two aluminized Mylar sheets are placed to minimize light leakage from the grooves. The scintillation light signal is captured by 1-mm-diameter multi-cladding Kuraray Y-11 (200) WLS fibers \cite{kuraray-catalogue}.
Each WLS fiber is coupled to a
Hamamatsu S13360-1325PE multi-pixel photon counter (MPPC) with $1.3 \times 1.3$ mm$^{2}$ active area. 
Its nominal photodetection efficiency (PDE) is 25\% at 450 nm, with a typical dark count rate of 70 kcps and 1\% crosstalk probability at 5 V over-voltage \cite{hamamatsu:mppc}.
32 MPPCs are organized in 4 rows and 8 columns with a pitch of 10~mm 
and mounted on a specially designed printed circuit board (PCB), as shown in Fig.~\ref{fig:prototype-mppc-pcb}. 

To provide a good coupling between MPPCs and WLS fibers, we designed an aluminum support plate with holes. The WLS fibers are glued inside the holes with EJ-500 epoxy glue \cite{EJ-500-glue}. 
Once the glue was cured, the WLS fiber ends were polished by a five-axis CNC milling machine with diamond cutters to ensure high-quality polishing. Fig.~\ref{fig:prototype-alu-plate} shows the WLS fibers glued to the aluminum plate, while Fig.~\ref{fig:prototype-fiber-end} highlights the quality of the polishing.
After this step, we mounted the \mppcpcb on the aluminum support plate.
Using this configuration, both the alignment and the distance between the MPPC and the fiber could be ensured with machining precision to provide response uniformity among the different channels.
Additionally, alignment pins were used.
Moreover, the EJ-500 glue was found to adhere efficiently to the aluminum, providing enough strength to resist the stresses and vibrations possibly induced by the CNC machining.
To prevent applying any pressure on the MPPC during the mounting of the \mppcpcb, a 1~mm spacer was inserted between the PCB and the aluminum plate. Given that the thickness of the MPPC is 0.95~mm, this configuration creates an air gap of 0.05 mm.
Thus, while we opted for the safer approach,
optical simulations show that this can reduce the light by up to 30\%, compared to the case of perfect contact.
If a higher light yield is required, it is common practice to place optical grease on the surface between the WLS fiber and the MPPC to maximize light transmission.
The back side of the \mppcpcb has a Samtec LSHM connector, connected to the front-end board via two HLCD flat cables.
These cables transmit the MPPC signals and supply the bias voltage.

The digitisation of the MPPC analogue signals is ensured by a dedicated front-end board (FEB), of the same type as used in the SuperFGD detector of the T2K experiment \cite{T2K:2026zms}.
The FEB is structured around eight CITIROC chips (Cherenkov Imaging Telescope Integrated Read Out Chip \cite{citiroc}).
The MPPC analogue high-gain signal was sampled by the peak
detector system, which means that the measured charge was obtained from the amplitude of the highest signal peak. 
A GPIO board is used as the master clock board and interface for the data-acquisition system. 
The whole setup is placed in a dark box to efficiently shield against environmental light.

\subsection{Calibration and event selection}

\begin{figure}[htbp]
    \centering
    \raisebox{+0.\height}{
    \begin{subfigure}{0.45\textwidth}
        \centering
        \includegraphics[width=\textwidth]{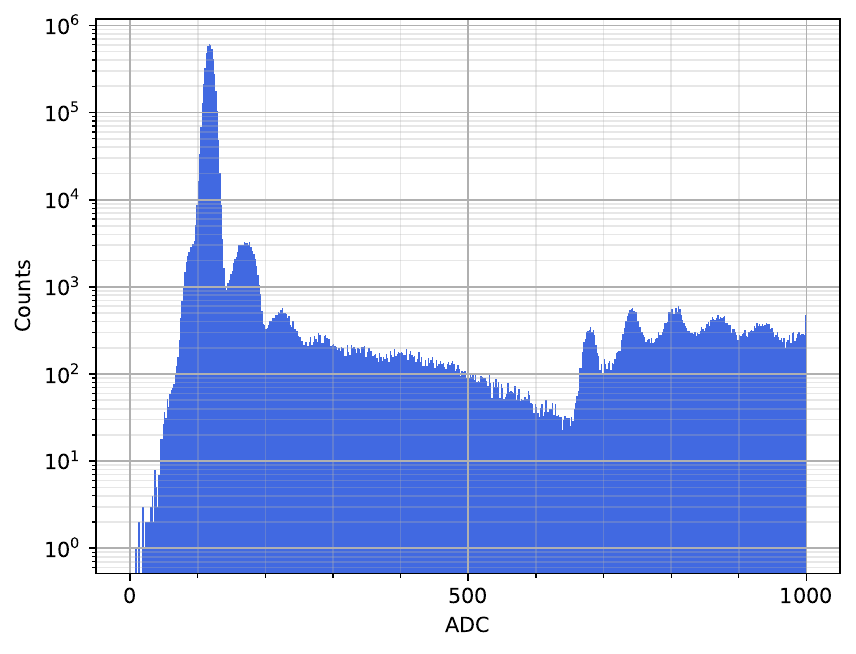}
        \caption{ADC distribution for a single readout channel.}
        \label{fig:calib-adc}
    \end{subfigure}
    }
    \hfill
    \raisebox{-0.07\height}{
    \begin{subfigure}{0.45\textwidth}
        \centering
        \includegraphics[width=\textwidth]{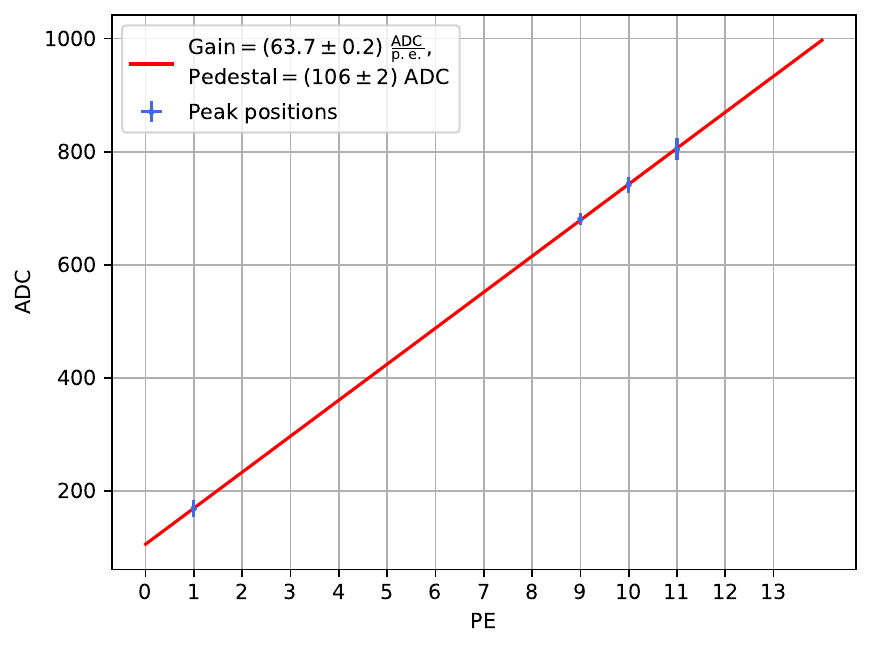}
        \caption{
        ADC to number of PE calibration fit.
        The pedestal and gain values are shown in the legend.
        }
        \label{fig:calib-gain}
    \end{subfigure}
    }
    \caption{
    Calibration fit of the number of measured ADC versus PE using a sample of cosmic rays for an example channel.
    }
    \label{fig:example_calib}
\end{figure}

The MPPC signal is digitized by the FEB in analog-to-digital count (ADC) units. To convert it into the number of photoelectrons (PE), calibration is performed separately for each readout channel with a sample of cosmic-ray data, looking at the lower part of the light yield spectrum.
Fig.~\ref{fig:calib-adc} shows an example light yield distribution used for identifying the electronics pedestal (first peak at low ADC) and for determining the gain, which is the ADC interval corresponding to adding one more PE.
The gain is computed by subtracting the pedestal from the recorded ADC distribution and then taking the distance between two consecutive peaks, each corresponding to a different number of PE. 
The position of the PE peaks has been computed with a Gaussian fit. 
The most pronounced PE peaks used for the calibration are those induced by %electronics hits from 
optical crosstalk (below 300 ADC) and low light yield cosmic events (above 600 ADC). Both peaks were used.
As shown in Fig.~\ref{fig:calib-gain} for a single channel, the slope of a linear fit of the ADC values as a function of the number of PE returns the gain.

\subsection{Light yield measurement}
\label{sec:measurements-light-yield}

Vertically through-going cosmic-ray particles are selected. 
On each readout view, the activated electronics channel (hit) with the highest number of PE is selected in each of the three \gluedcubelayer{s}. 
Three hits are required to align along the same column on each side view, with a minimum of 3 PE. 
An example of a selected event is shown in Fig.~\ref{fig:prototype-event}.
To measure the light yield and the optical crosstalk, we use solely the hits from the central \gluedcubelayer, while the top and bottom layers are used for the event trigger and selection. 
This allows us to minimize the spread in angle and, thus, the path length in the scintillator cube.
The light yield was measured with three \mppcpcb{s} each on a different orthogonal view, to provide the full 3D track of the cosmic ray, as shown in Fig.~\ref{fig:prototype-setup}. 
On each view, different sets of channels were tested.

\begin{figure}[htbp]
    \begin{subfigure}{0.49\textwidth}
        \centering
        \includegraphics[width=\textwidth]{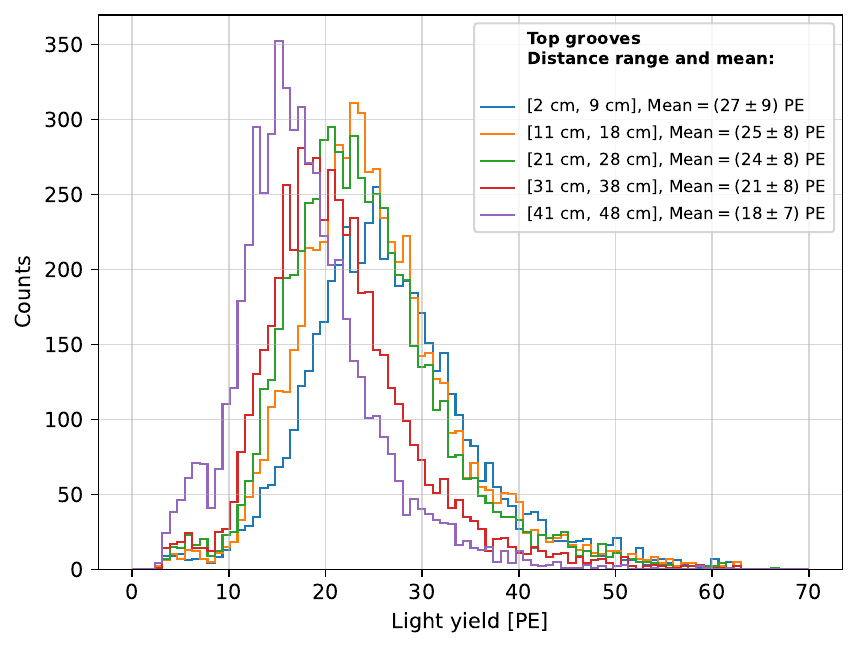}
        \caption{Side view top groove: light yield for \mppcpcb{s} at different positions.}
        \label{fig:light-yield-sideview-topgroove}
    \end{subfigure}
    \hfill
    \begin{subfigure}{0.49\textwidth}
        \centering
        \includegraphics[width=\textwidth]{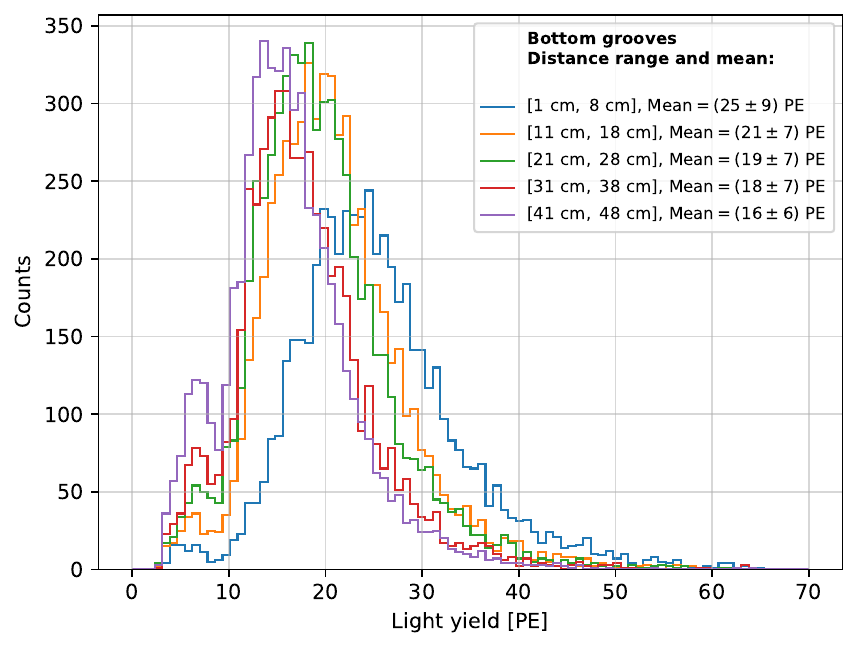}
        \caption{Side view bottom groove: light yield for \mppcpcb{s} at different positions.}
        \label{fig:light-yield-sideview-bottomgroove}
    \end{subfigure} \\
    \centering
    \raisebox{-0.08\height}{    
    \begin{subfigure}{0.45\textwidth}
       \includegraphics[width=\textwidth]{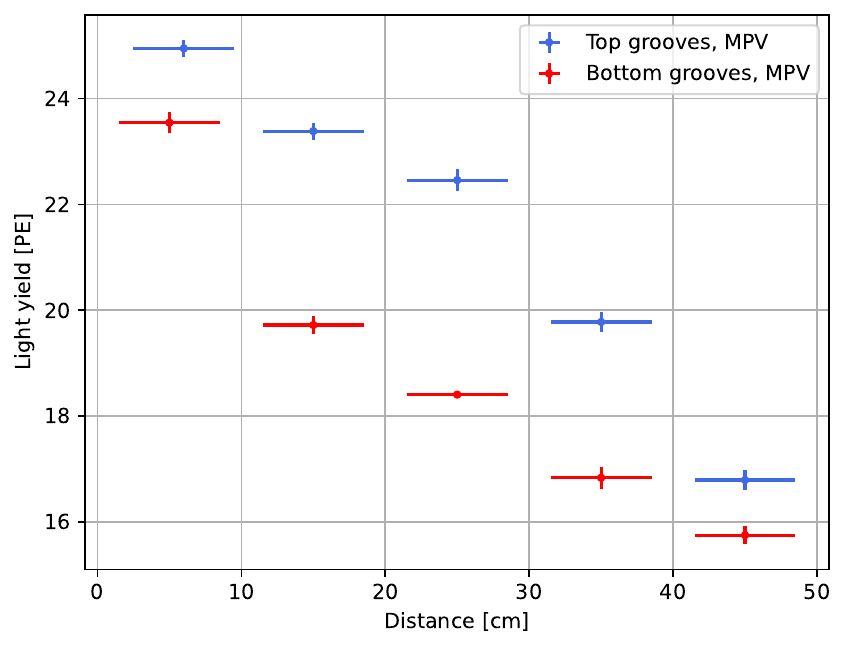}
        \caption{
        Light yield versus distance from the MPPC. The MPV of the distributions in Fig.~\ref{fig:light-yield-sideview-topgroove} (blue) and \ref{fig:light-yield-sideview-bottomgroove} (red) is plotted. 
        }
        \label{fig:light-yield-sideview-vs-distance}
    \end{subfigure}}
    \hfill
    \begin{subfigure}{0.45\textwidth}
        \centering
        \raisebox{-0.5\height}{\includegraphics[width=\textwidth]{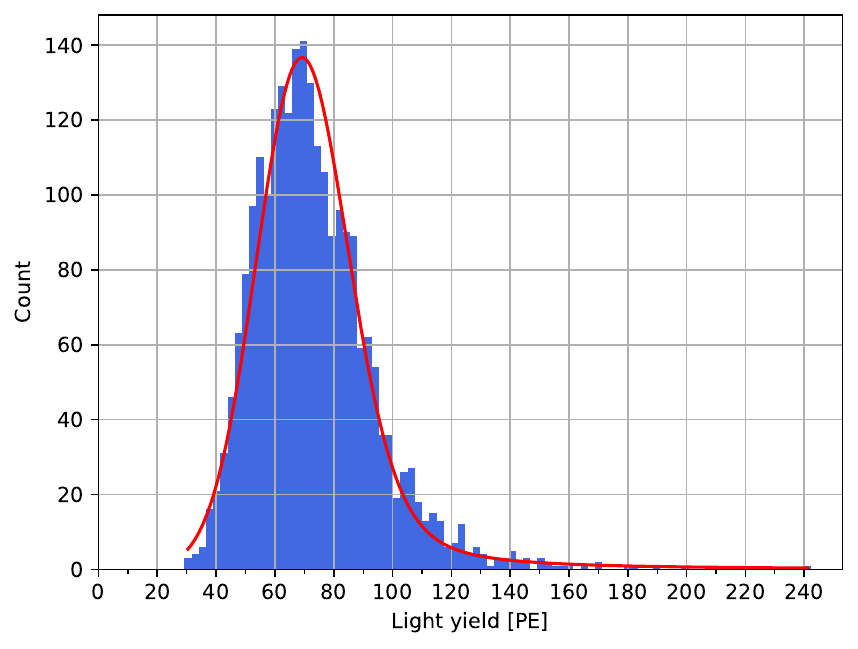}}
        \caption{
        Light yield of a top-view channel for a \mppcpcb at the centre of the \gluedcubelayer.
        }
        \label{fig:light-yield-topview}
    \end{subfigure}
    \caption{Light yield distributions in units of PE averaged over the channels of a \mppcpcb.
   }
    \label{fig:light-yield}
\end{figure}

In Figs.~\ref{fig:light-yield-sideview-topgroove} and~\ref{fig:light-yield-sideview-bottomgroove} the light yield distributions for the side views averaged over the MPPCs in a single \mppcpcb are shown for different groups of channels. 
The most probable values (MPV) are extracted from the fit of a Landau-Gaussian convolution function to each distribution and plotted in Fig.~\ref{fig:light-yield-sideview-vs-distance} as a function of the distance between the selected cosmic ray trajectory and the MPPC.
Here, the points correspond to the MPV of the histograms in Figs.~\ref{fig:light-yield-sideview-topgroove} (blue) and 
\ref{fig:light-yield-sideview-bottomgroove} (red). 
While the horizontal bars indicate the size of the \mppcpcb,
the vertical bars represent the uncertainty on the peak position from the Landau-Gaussian convolution fit. 
The MPV light yield varies between $23.5$--$25.0$~PE per channel at the position closest to the 
MPPC
and drops to approximately $15.8$--$16.6$~PE at the farthest position from the MPPC, driven by the light attenuation in the WLS fiber. 
One can note a systematic difference in the light yield of 1--3~PE between the two side views, that is, between WLS fibers in the top and bottom grooves. 
Visual inspections did not spot any particular difference in the quality of the groove polishing. 
On the other hand, one possible explanation may be the different way in which a WLS fiber sits in each groove: due to gravity, in the top groove, the fiber remains mainly in contact with the plastic scintillator; 
instead, in the bottom groove, the fiber could touch the Mylar sheet and be separated from the scintillator, creating an air gap.
This may reduce the fiber light trapping efficiency. 
Potentially, the light yield could be increased and made more uniform across the channels by depositing optical grease in the grooves. 
We defer these studies to future work.

Unlike the side views, the WLS fibers of the top view are hosted in holes and not grooves.
On the other hand, since cosmic rays are mainly vertical, the light yield from the top view is expected, to the first order, to correspond to the sum of scintillation in three consecutive cubes. 
This feature can be observed in Fig.~\ref{fig:light-yield-topview}.
The light attenuation in the WLS fiber is small, due to the distance between the MPPC and the cosmic-ray trajectory being constrained to between 1 and 3 cm.  
The light yield (MPV), around 73~PE,
is consistent with three times the side-view light yield of a single layer.

\begin{figure}[htbp]
    \centering
    \begin{subfigure}{0.45\textwidth}
        \centering
        \includegraphics[width=\textwidth]{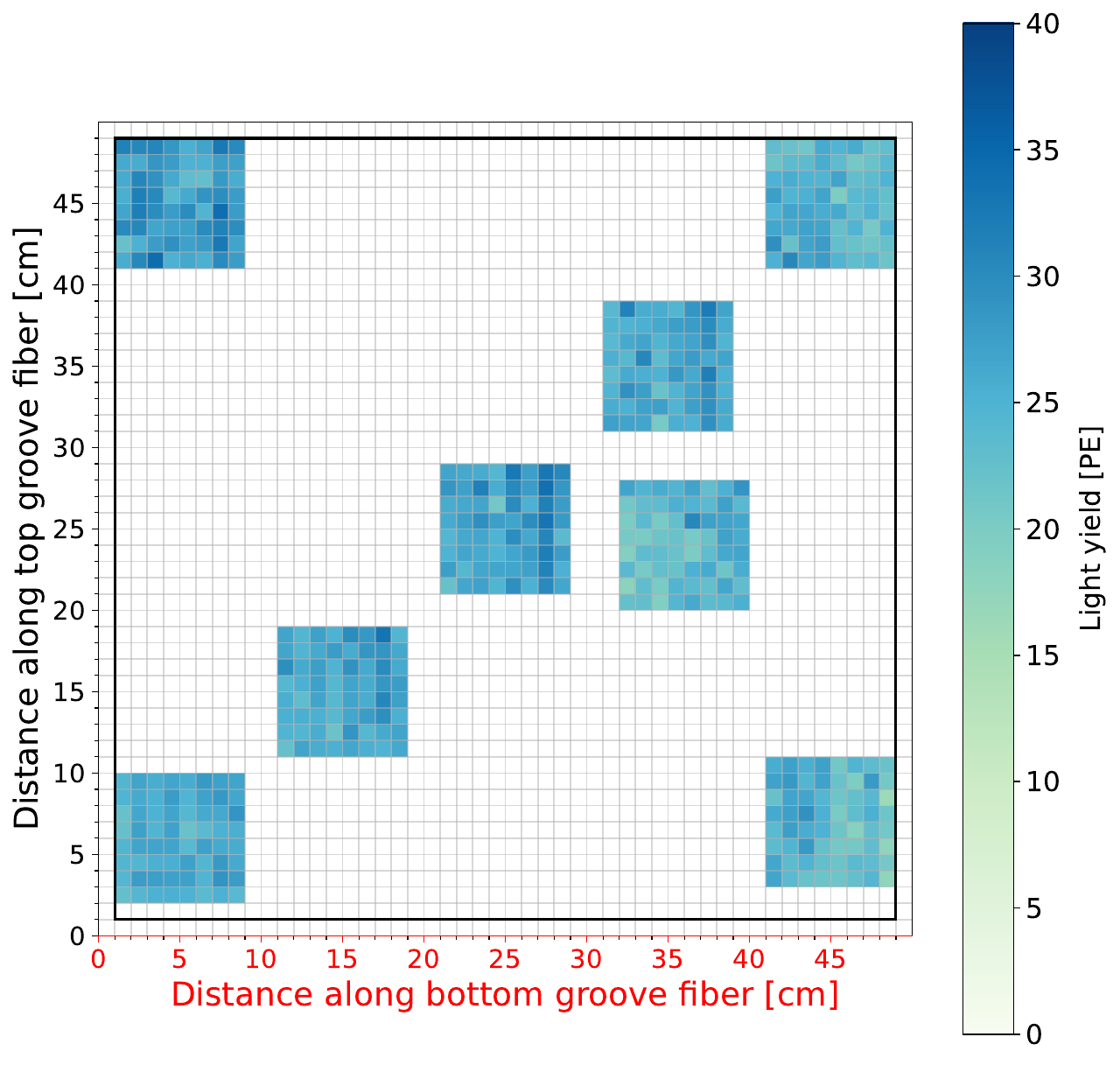}
        \caption{Side view - top groove .}
    \end{subfigure}
    \hfill
    \begin{subfigure}{0.45\textwidth}
        \centering
        \includegraphics[width=\textwidth]{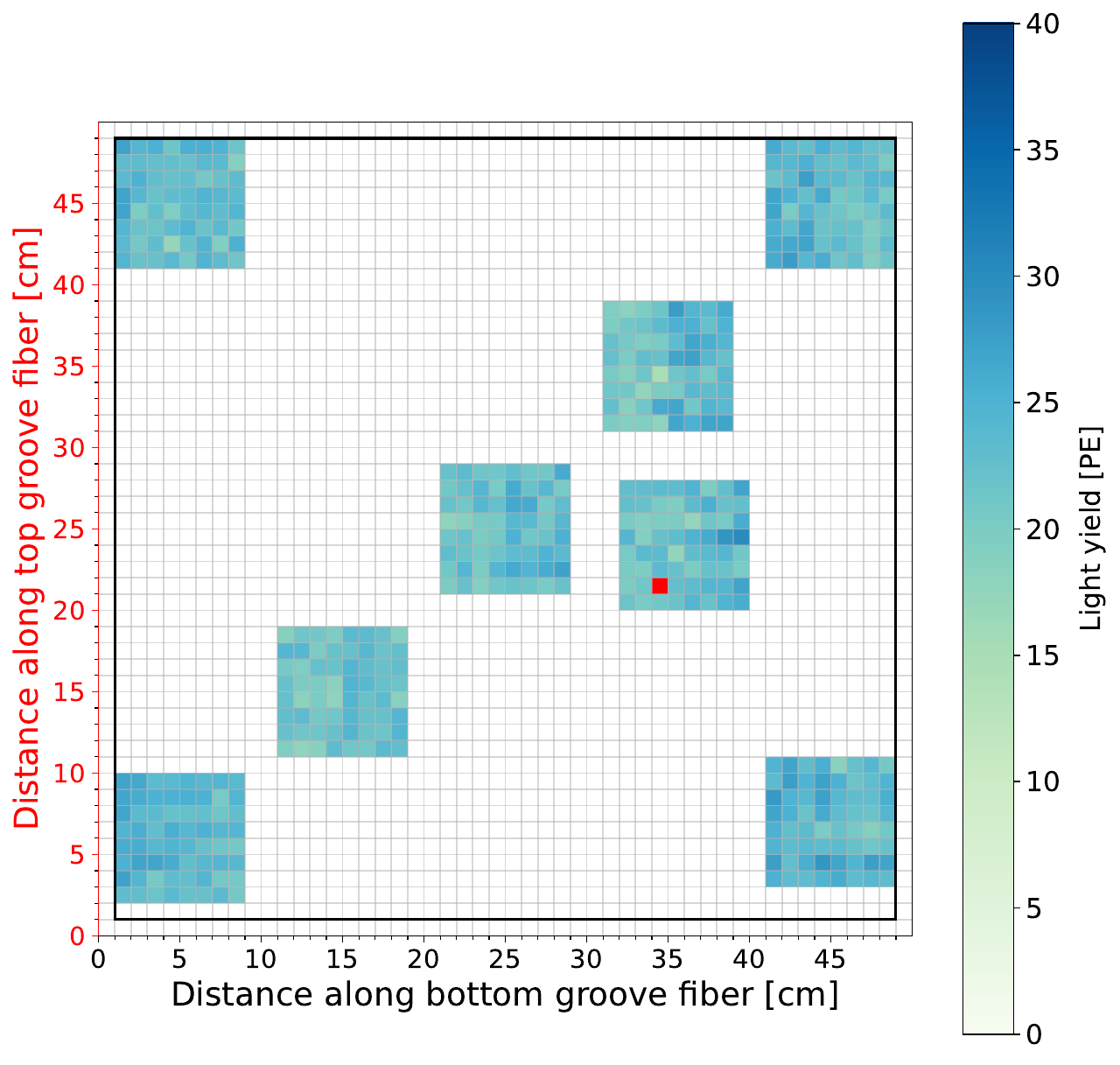}
        \caption{Side view - bottom groove.}
    \end{subfigure}
    \caption{Light yield corrected for the attenuation length for the middle \gluedcubelayer measured at different cube positions. 
    The colored regions show where the measurements were performed. 
    The axis marked with red color indicates the side where \mppcpcb was mounted. 
    }
    \label{fig:ly_uniform_corr_percube}
\end{figure}

To check the response uniformity across the \gluedcubelayer, the light yield measured from different cubes was compared.
Since cosmic rays are mainly vertically directed, the channels on the side views allowed for the selection of a large statistical sample where the energy deposited in a single cube could be separated from that of the others. 
Given the size of the \gluedcubelayer, the light yield was corrected for the attenuation length in the fiber.
The light yield as a function of the distance from the MPPC was fitted with a
single 
exponential function. 
In Fig.~\ref{fig:ly_uniform_corr_percube}, the light yield as a function of the cube position is shown for the middle layer for the top and bottom grooves.
Although not all the cubes could be tested, results indicate a rather uniform response across the \gluedcubelayer.
A difference in the light yield between the top and the bottom grooves is observed, as already highlighted in Fig.~\ref{fig:light-yield}.
Anyhow, the remaining cube-to-cube non-uniformity can be calibrated if more statistics are available. In fact, the three-dimensional granularity, as well as the three readout views, allows for the selection of single cubes for single-track events and the independent measurement of the light yield.

\subsection{Optical crosstalk measurement}
\label{sec:measurements-cross-talk}

In this work, the crosstalk is defined as the ratio between 
the light yield measured by the adjacent WLS fibers  
and that measured from the fiber crossed by the particle. 
The definition follows that used in Ref.~\cite{Boyarintsev:2021uyw}. 
It was measured from the same sample of cosmic muons used for the light yield measurement in Sec.~\ref{sec:measurements-light-yield}.

\begin{figure}[htbp]
    \centering
    \begin{subfigure}{0.45\textwidth}
        \centering
        \includegraphics[width=\textwidth]{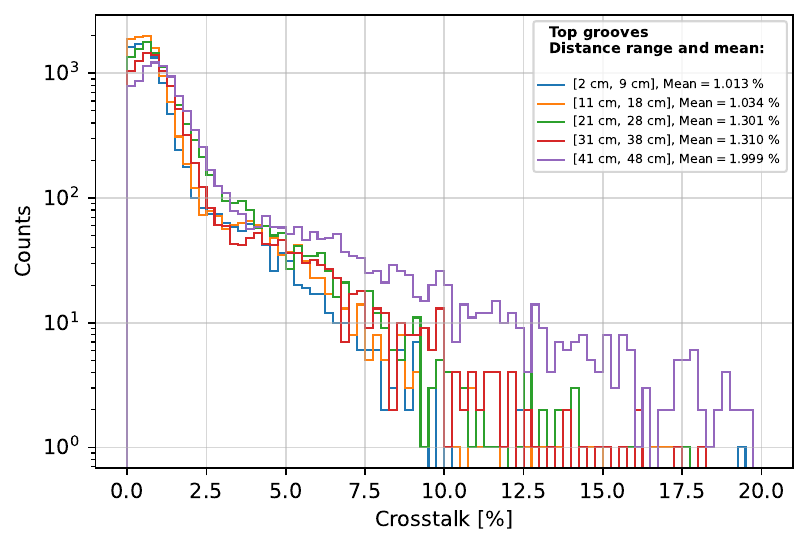}
        \caption{Top groove view.}
        \label{fig:xtalk_dis_view_a}
    \end{subfigure}
    \hfill
    \begin{subfigure}{0.45\textwidth}
        \centering
        \includegraphics[width=\textwidth]{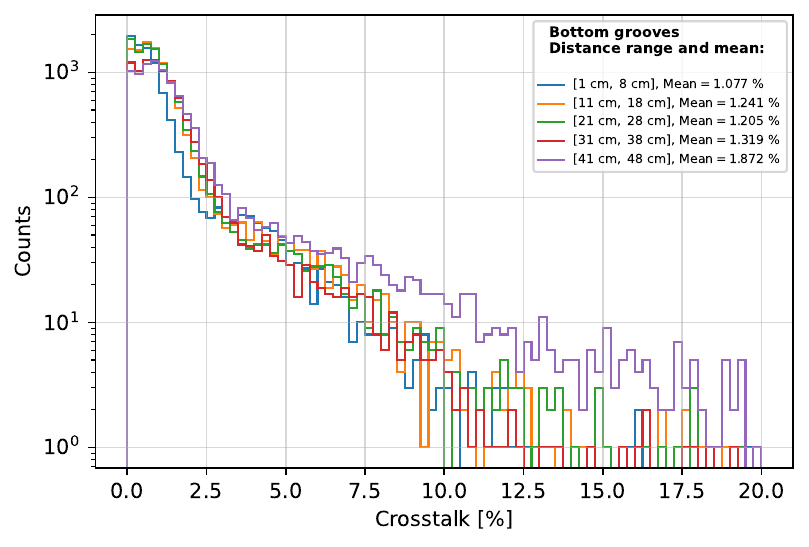}
        \caption{Bottom groove view.}
        \label{fig:xtalk_dis_view_b}
    \end{subfigure} \\
    \centering
    \begin{subfigure}{0.45\textwidth}
    \includegraphics[width=\textwidth]{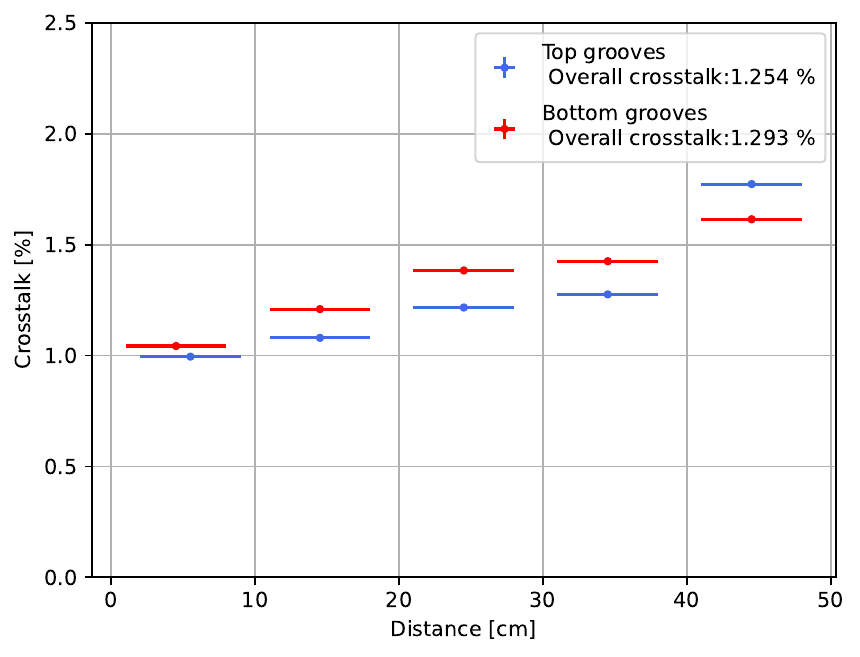}
    \caption{Crosstalk fraction as a function of the MPPC to cosmic muon distance for the side view top (blue) and bottom (red) grooves.
    The horizontal error bar represents the size of the \mppcpcb.
    }
    \label{fig:xtalk_vs_distance}
    \end{subfigure}
    \caption{
    Optical crosstalk measured from a sample of cosmic-ray muons at different positions in the glued-cube layer, averaged over the channels of a single MPPC-PCB. Panels \subref{fig:xtalk_dis_view_a} and \subref{fig:xtalk_dis_view_b}
    show the event-by-event distribution of the crosstalk fraction (light yield in the adjacent fiber over light yield in the fiber crossed by the particle), binned by distance from the MPPC. 
    Panel \subref{fig:xtalk_vs_distance} shows a distance-binned, light-yield-weighted ratio pooled over all events, which is not the ratio of means of panels \subref{fig:xtalk_dis_view_a} and \subref{fig:xtalk_dis_view_b} (see discussion in text).
    }
    \label{fig:xtalk}
\end{figure}

Fig.~\ref{fig:xtalk} summarizes the results of the optical crosstalk measurements between cubes in different parts of the \gluedcubelayer. 
Figs.~\ref{fig:xtalk_dis_view_a} and \ref{fig:xtalk_dis_view_b} show, for each MPPC-distance bin, the event-by-event distribution of the ratio between the light yield in the adjacent fiber and in the fiber crossed by the particle, for the top- and bottom-groove views; 
the long tail at large distances 
is due to
the occasional presence of delta-rays, that can cross the boundary and deposit energy in the adjacent cubes,
as well as to the growing relative impact of pedestal fluctuations. 
In fact, these distributions exhibit a tail that extends to larger values as the distance from the MPPC increases; this reflects events in which the light yield in the crossed fiber is small due, for example, to attenuation or a shorter path of the particle within the cube, so that even a modest pedestal fluctuation in the neighboring channel produces an anomalously large ratio for that single event.
Hence, the arithmetic mean of the per-event distributions, quoted in the legends of Figs.~\ref{fig:xtalk_dis_view_a} and \ref{fig:xtalk_dis_view_b}, is not a robust estimator of the crosstalk fraction.
The \textit{crosstalk fraction} quoted in Fig.~\ref{fig:xtalk_vs_distance} and used in the following discussion is instead computed as the ratio of the light yield summed over all events in the adjacent fiber to the light yield summed over all events in the crossed fiber, within each distance bin. This pooled estimator effectively weights each event by its own light yield, so that the low-signal, pedestal-dominated events that inflate the per-event mean contribute little to either sum, making it considerably more stable against statistical fluctuations than the simple mean.
The same approach is discussed and used in Ref.~\cite{Li:2025dff}.

We find that the crosstalk fraction is always below 2\%,
consistent with Ref.~\cite{Boyarintsev:2021uyw}.
Thus, the chance of finding a particle track wider than the single sensitive cube is very small. This is particularly important in high-multiplicity events such as TeV neutrino interactions at FASER.

As a side note, since cosmic muons are directed vertically, the crosstalk was computed only between cubes of the middle \gluedcubelayer.
On the other hand, the crosstalk between consecutive \gluedcubelayer{s} is attributed to the quality of the reflective Mylar sheet rather than to the \gluedcubelayer itself. 
For instance, multiple sheets or higher-performance reflectors 
such as the white diffuser lumirror \cite{lumirror,6168236}
or specular mirror  \cite{3m_df2000ma,Li:2025dff}
sheets may also be used.
Moreover, it is worth noting that the quality of the white paint on the top and bottom faces of the \gluedcubelayer serves to minimize the cube-to-cube crosstalk within the same layer by avoiding leaks due to reflections from the specular Mylar sheet.
Thus, the results reported in this work are sufficient to certify the quality of the optical separation within the single \gluedcubelayer.

\section{Conclusions}
\label{sec:conclusions}

In this work we demonstrated the scalability of the \gluedcubelayer technology to the large sizes required for future particle detectors and calorimeters.
It consists of a monolithic layer of $48 \times 48$ optically-separated plastic scintillator cubes with 9 mm edges, glued together with machining precision and read out by three orthogonal WLS fibers to provide three projections of the particle interaction.
The specific size of the \gluedcubelayer was chosen to fit the preliminary detector design for the possible future upgrade of the FASER neutrino detector \cite{FASER:2025myb}.

To prove the scalability to large-size detectors, we first assessed the geometrical tolerances, as well as the flatness, and certified its suitability through metrological studies and mechanical tests.

Then, we characterized the optical performance in terms of light yield and crosstalk, obtaining satisfactory results.
One may note that while the optical crosstalk is comparable to that reported in Ref.~\cite{Boyarintsev:2021uyw} for the $5 \times 5$ cube prototype, the light yield is lower by almost a factor of two.
The reasons for such a difference can be multiple: 
most importantly, the use of MPPCs with a lower PDE, i.e., 25\% for S13360-1325PE in this work, versus 40\% for S13360-1350CS;
the five times longer WLS fibers, thus a stronger attenuation of the light;
the air gap present between the MPPC and the WLS fiber enforced by the spacers on the \mppcpcb.
While the points listed above would suffice in explaining the difference in light yield, 
additional disparities between the prototypes may potentially occur in the MPPC bias voltage and the transparency of the WLS fibers grooves. 
None of the features above can be attributed to the considerably larger size of the \gluedcubelayer.
On the other hand, for applications that require a higher light yield, one can apply: 
optical grease between the WLS fiber and the MPPC and/or in the groove hosting the WLS fibers; 
mirror paint at the end of the WLS fiber;
or increase the MPPC bias voltage, if the experimental condition can accept a higher dark count rate.

While the \gluedcubelayer has been proven, efforts are continuing to further improve the technology,
aiming to: 
reduce the thickness of the reflective epoxy glue from 1 mm to 0.6 mm while ensuring sufficient rigidity for the \gluedcubelayer without sacrificing optical separation;
producing cubes of smaller size;
and integrating an absorber layer, made out of tungsten powder embedded in a polymer matrix, 
to the \gluedcubelayer. 
Moreover, work is also underway to modify the scintillator.
The maximum luminescence of the scintillating polystyrene can be shifted from 418 nm to 430 nm to better match the absorption spectrum of the Kuraray Y11 WLS fibers and potentially increase the detector's light yield.
Developments of an improved radiation-resistant polystyrene scintillator layer with a peak luminescence around 530 nm are also underway.

Finally, it is worth noting that sizes up to or beyond  $1 \times 1~\text{m}^2$ are also within reach and do not pose major technical limitations.

\section{Acknowledgements}

This work was supported by the grants 
20FL21\_232700 and PCEFP2\_203261
from the Swiss National Science Foundation.
We thank the members of the FASERCAL working group for discussions of the results, especially Charlotte Cavanagh for her comments.

\subsection{Competing interests}
The authors declare no competing interests.

\bibliographystyle{utphys}
\bibliography{GluedCubes.bib}

\iffalse

\fi

\end{document}